\documentclass[11pt,a4paper,dvipdfmx]{article}
\usepackage{aer}
\usepackage{aer-osborn}
\usepackage{graphicx}
\usepackage[]{natbib}
\usepackage{url}

\begin{document}

\title{\Large\bf
  An Analysis of the Japanese Credit Network}

\def\addnum#1{$^#1$}

\author{\sl
  G. De Masi   \addnum1,
  Y. Fujiwara  \addnum2{$^\dag$},
  M. Gallegati \addnum1,\\\sl
  B. Greenwald \addnum3,
  J. E. Stiglitz \addnum3
  \\[10pt]
  {\sl\small\addnum1
    Universita Politecnica delle Marche, Ancona, Italy}\\
  {\sl\small\addnum2
    ATR Laboratories, Kyoto, Japan}\\
  {\sl\small\addnum3
    Columbia University, New York, USA}}
    
\date{}
\maketitle

\begin{abstract}
  An analysis of the Japanese credit market in 2004 between banks and
  quoted firms is done in this paper using the tools of the networks
  theory. It can be pointed out that: (i) a backbone of the credit
  channel emerges, where some links play a crucial role; (ii) big
  banks privilege long-term contracts; the ``minimal spanning trees''
  (iii) disclose a highly hierarchical backbone, where the central
  positions are occupied by the largest banks, and emphasize (iv) a
  strong geographical characterization, while (v) the clusters of
  firms do not have specific common properties. Moreover, (vi) while
  larger firms have multiple lending in large, (vii) the demand for
  credit (long vs. short term debt and multi-credit lines) of firms
  with similar sizes is very heterogeneous.
\end{abstract}

\vskip2cm



\vskip2cm
\noindent\hrulefill

\noindent
Correspondence:
  Yoshi Fujiwara,
  ATR Laboratories, MIS,
  Kyoto 619-0288, Japan. \\
  tel: {\tt +81-774-95-1404} / fax: {\tt +81-774-95-1409} /
  email: {\tt yoshi.fujiwara@gmail.com}


\noindent
{$^\dag$} Also affliated with:
  Kyoto University, Kyoto, Japan \\[20pt]

\newpage
\section{Introduction}

Debt-credit relationships between firms and banks have a long history
in economics \citep{schumpeter1911ted}.
It has been widely recognized since \citet{debreu1959tv} that
integrating money in the theory of value represented by the General Equilibrium model
is problematic at best. No economic agent can individually decide to
monetize alone; monetary trade should be the equilibrium outcome of
market interactions among optimizing agents. The use of money --- 
a common medium of exchange and a store of value --- implies that
one party to a transaction gives up something valuable (for instance,
his endowment or production) for something inherently useless (a
fiduciary token for which he has no immediate use) in the hope of
advantageously re-trading it in the future. Since credit makes sense
only if agents can sign contracts in which one side promises future
delivery of goods or services to the other side,
markets for debt in equilibrium are meaningless. 
A non-mainstream approach to cope with the financial
stability based on the understanding of debt-credit relationships
between heterogeneous interacting agents is strongly required.

This point of heterogeneity is linked to the existence of some
underlying autocatalytic process at a lower level of the system. An autocatalytic
process is a dynamic process in which the growth of a quantity is to
some extent self-perpetuating, as in the case when it is proportional
to its initial value. The existence of an autocatalytic process
implies that looking at the average, or most probable, behavior of the
constituent units is non-representative of the dynamics of the system:
``autocatalyticity insures that the behavior of the entire system is
dominated by the elements with the highest auto-catalytic growth rate
rather than by the typical or average element'' \citep{solomon2007cr}.
In the presence of autocatalytic processes, even a small amount of
individual heterogeneity invalidates any description of the behavior
of the system in terms of its ``average'' element: the real world is
controlled as much by the {\it tails\/} of distributions as by means
or averages. We need to free ourselves from {\it average\/} thinking
(\citet{anderson1997std} in the context of statistical physics;
\citet{brock1999ser} in that of economics; see also
\citet{mantegna2000iec}).

The purpose of this paper is to investigate the structure and
heterogeneity of debt-credit relationships by applying the recently
developed tools of {\it network analysis\/} to an empirical data of
Japanese debt-credit network to a nation-wide scale. This
investigation is quite relevant to financial stability issues; for
instance, the failure of a firm heavily indebted with a bank may
produce important consequences on the balance sheet, or the financial
status, of the bank itself. If a bank's supply of credit is depleted,
total supply of loans is negatively affected and/or the rate of
interest increases, thus transferring the adverse shock to the other
firms. Therefore, the study of structure of the links and their weights
allows to gain some insights in the financial stability of the
economic system and to develop new economic policy tools.

Let us first briefly review the literature mainly based on traditional
methodologies. The exploration of the structure of credit relationships
among banks and firms recently acquired increasing importance. The
availability of new and large data sets allowed researchers to analyze
the number of credit relationships between firms and banks in
different years and countries (see %
\citeauthor{diamond1984fia},~\citeyear{diamond1984fia},
\citeauthor{ongena2000dnb},~\citeyear{ongena2000dnb}, for example).
These studies show that, except for a few cases of very cash-rich
firms, internal financing is only limited; short-term and long-term
loans play a crucial role in the investment expenditure of the
economic system in most developed countries.

An important aspect is the empirical analysis on the single
{\it vs.}~multiple banks-firms credit relationships (%
\citeauthor{agarwal2001bfr},~\citeyear{agarwal2001bfr};
\citeauthor{farinha2002ssm},~\citeyear{farinha2002ssm}; and
\citeauthor{ogawa2007jfp},~\citeyear{ogawa2007jfp}%
) that are based on cross-country comparisons. One can observe the
presence of two paradigmatic examples and many intermediate cases,
i.e. the {\it bank oriented\/} example of Germany, Italy and Japan
(e.g. less than 3\% of Italian firms have single bank relationships),
characterized by a close firm-bank relationship, and the {\it market
  oriented\/} paradigm of the Anglo-Saxon system (e.g., in the UK,
25\% of firms maintain only one bank relationship).
Other countries, such as the EU ones, are in the
middle range between these two cases.

In institution-oriented countries, quite often a single firm may be
influenced by the so-called {\it inside bank}. In these cases, the
inside bank has a more favorable access to information about the
actual financial condition of a particular firm. In the literature, a
firm is defined as {\it bank influenced\/} when a particular bank owns
more than 50\% of the firm's share or if the chair of the supervisory
board is a banker \citep{agarwal2001bfr}. Even in Germany and Japan,
where the main bank often plays a dominant role, firms subscribe loans
contracts with other banks. Moreover, \citet{sterken2002dnb} states
that the presence of a credit line with a main bank attracts more
loans from other banks, signaling an asymmetric information problem.

Indeed, the theory of the optimal number of bank relationships gave
gave a number of advantages and disadvantages in the choice of single
and/or multiple relationships. On the firm side, a single bank
relationship comes from the minimization of costs in transactions and
monitoring, while the firm could benefit in a competing market of
banks; this implies a growth in the number of relationships.  Multiple
banks lending guarantees the firm against the risk of liquidation. On
the bank side, financing firms with multiple bank relationships allows
to pool the risk of failure of firms. Single linkages, on the other
hand, would give the bank greater control on the financial choices of
the firm.

Moreover the tendency to multiple or single relationships changes in
time, varying with internal and external conditions. There is
evidences that some firms start with a single relationship and after
some time they switch to multiple links in conditions of growth
opportunities \citep{farinha2002ssm}. In particular conditions of
financial distress, an evolution of the structure of the lending
relationships can be observed: for instance, in Japan, during the
bubble period, firms tended to rely on a single relationship
\citep{ogawa2007jfp}.

Now let us turn our attention to the Japanese system of banks-firms
credit relationships. In the presence of {\it keiretsu}, a terminology
for industrial corporate groups, firms have a strong and long-lasting
relationship with the so-called {\it main bank} (see
\citet{aoki1994jbs} for a review). The firm is particularly dependent
on the main bank for financing because of the information advantage
over other potential lenders; this is particularly evident in
conditions of financial turbulence \citep{spiegel2003fta}.
Bank-influenced firms should enjoy increased access to capital through
easier access to bank debt or preferential terms on loans, but on the
other hand there may be some negative effects. Close relationships
allow banks to have a major role in the corporate governance
structure, like the representation of the bank on the firm's
supervisory board. Banks that handle the majority of new equity issues
of the firm often place them among their customers, but on the other
hand, they can influence the financial decisions of firms: in fact, in
the case of firm distress, they can force firms to issue equity to pay
bank debt.

The Japanese system is characterized by the presence of different
types of banks: long-term credit banks, trust banks, city banks,
regional banks, secondary regional banks and insurance companies. In
particular long-term credit banks do not have affiliation with
corporate groups and in-house credit analysis; trust banks have
long-term credit to designed sectors and supplement city banks. In
Japanese development after the Second World War, long-term credit
banks played a crucial role because the financing of firms by issuing
bonds had been strictly regulated (only after 1985 a few big firms
were able to issue bonds). {\it Institution-oriented\/} markets have
been extensively investigated. While in Germany banks-influenced firms
do benefit from increased access to capital (but there is no evidence
to support the hypothesis of either profitability or growth) and the
payment of interest rate to debt ratio is higher for them
\citep{agarwal2001bfr}, in Japan, banks influence firms to decide
about low-risk investment decisions (lower debt-equity ratio) and
bank-related firms are less profitable than other ones
\citep{ogawa2007jfp}. \citet{ogawa2007jfp} carried out a detailed
analysis of dependency of the number of long-term credit relationships
on the characteristics of the firm; they emphasized that, while the
largest firms have the largest number of banks relationships, the
number of relationships is strongly positively correlated with
solvability and R\&D and inversely with the liquidity of the firm. A
higher profitability (ROA), debt-on-asset ratio (DAR) and lower
liquidity (LAR) lead to more banking relationships according to
\citet{sterken2002dnb}, while \citet{kano2006ivb}, in studying the
small and medium enterprises, noticed that they benefit most from
bank-borrower relationship when they do not have audited financial
statements and when they borrow from small banks in less competitive
markets.

Since 1992, the Japanese banking system has experienced a sizable
deterioration in its financial conditions \citep{brewer2003brd}.
Commercial banks have recorded cumulative loan losses of about 83
trillion yen. These losses reduced the bank capitalization and led to
the failure of three large banks (and other small banks). The very
poor financial conditions of those major banks affected the whole
credit system, especially those in weaker financial conditions. In
order to increase the financial stability of the system, in 1997 the
Japanese regulators liquidated a large city bank and nationalized 2 of
the 3 largest long-term credit banks
(\citeauthor{brewer2003brd},~\citeyear{brewer2003brd} where it
is emphasized that the banks' failure negatively affected the stock
prices of firms that had lending relationships with the failed banks).
In this paper, we analyze the bank-firm relationships in Japan in
2004, using the network theory, where network is the set of {\it
  nodes\/} (two types in our case: banks and firms) and {\it links\/}
(debt/credit contracts between them).

There are some recent investigations of the ``inter-bank market''
\citep{iori2007nai, boss2004nti}, but to the best of our knowledge
only a few studies of the financial credit market (see
\citet{masi2007bft} and \citet{fujiwara2009stc}). The advantages of
these network analyses include (a) characterization by statistical
features of a large-scale network (section~\ref{sec:nw_bf}), (b)
extraction of cohesive groups or communities in the network, and also
an extensive analysis to hierarchical structures
(section~\ref{sec:nw_banks}, \ref{sec:nw_firms}), and (c) correlation
of those statistical features
to some attributes of nodes/links (section~\ref{sec:financial_status}).
The point (c) here implies that the network analysis is just a
complementary methodology to those traditional ones that were briefly
reviewed in the above. The method can potentially provide scaffolding
which will enable one to build a model of the network under study and
a dynamics on it, that is, financial stability issues in our case.

The paper is organized as follows: section~\ref{sec:data} describes
our data on the multiple lending relationships in Japan.
The definition of the credit network is explained in detail
considering several topological measures in section~\ref{sec:nw},
and the methodology is applied to the dataset in order to describe
the architecture of the empirical network (section~\ref{sec:nw_bf}).
The following sections are, respectively, dedicated to an analysis of
the properties of the hierarchical clustering of co-financing banks,
to an investigation of the co-financed firms by sectors and to the
analysis of possible effects of the financial conditions on the
topology of credit relationships (sections~\ref{sec:nw_banks},
\ref{sec:nw_firms}, \ref{sec:financial_status}).
Section~\ref{sec:conclusion} concludes.

\section{The data set}\label{sec:data}

Our database is based on the survey of firms quoted in the Japanese
stock-exchange markets and on the financial statements publicly
reported by each quoted firm. The data were compiled by {\sl Nikkei
  Media Marketing, Inc.\/} and commercially available. The financial
statements and surveys include the information about each firm's
borrowing from financial institutions, the amounts of borrowing,
classified into short-term and long-term borrowings. ``Long-term''
borrowing is defined by scheduled repayment period exceeding one year,
and ``short-term'' borrowing refers to the other cases.

The financial institutions consist of {\sl long-term credit banks},
{\sl city banks}, {\sl regional banks\/} (primary and secondary), {\sl
  trust banks\/} and {\sl insurance companies}, basically all the
financial institutions in Japan, which we refer to as ``banks'' in
this paper. We also employ the database of the financial statements
for the banks except insurance companies. This database is
systematically compiled and maintained by the ``Japanese Banks
Association'' and is publicly available.

The numbers of banks and firms in the years 2000-2005 are reported in
table~\ref{tab:numbers}.

\begin{table}[htbp]
\centering
{\small
\begin{tabular}{|c|c|c|c|}
  \hline
  year & firms & banks & links \\
  \hline
  2000 & 2,629 & 211 & 27,389 \\
  2001 & 2,714 & 204 & 26,597 \\
  2002 & 2,739 & 202 & 24,555 \\
  2003 & 2,700 & 192 & 22,585 \\
  2004 & 2,701 & 190 & 21,919 \\
  2005 & 2,674 & 182 & 21,811 \\
  \hline
\end{tabular}
\caption{Numbers of firms, banks and links in our dataset.}
\label{tab:numbers}
}
\end{table}
 
Typical numbers of banks are, for instance in the fiscal year of 2004,
seven city banks (which include the four majors: the {\sl Bank of
  Tokyo-Mitsubishi}, the {\sl Mizuho Bank}, the {\sl Sumitomo-Mitsui
  Bank\/} and the {\sl UFJ Bank\/}), 64 regional banks, 48 secondary
regional banks and 9 trust banks; the rest are long-term credit
banks (which include two big banks: the {\sl Shinsei Bank\/} and the
{\sl Aozora Bank\/}) and insurance banks. As described previously, the
Japanese financial institutions experienced great damage in the
national financial system after the Bubble crash in 1990, and the
merger of banks restructured it. The process is still going on:
the {\sl Bank of Tokyo-Mitsubishi\/} and the {\sl
  UFJ Bank\/} have been merged into a single (and the largest)
financial group. The merger of banks is still going on since 2000
and can be emphasized by the decrease of the total number of banks.
This situation is common to all types of banks (see table~\ref{tab:banks}). The
other banks in each year are mostly life and non-life insurance banks
and a few government-affiliated financial institutions.

\begin{table}[htbp]
\centering
{\small
\begin{tabular}{|l|r|r|r|r|r|r|}
  \hline
  year & 2005 & 2004 & 2003 & 2002 & 2001 & 2000 \\
  \hline
  city & 7 & 7 & 7 & 7 & 9 & 9 \\
  regional & 64 & 64 & 64 & 64 & 64 & 64 \\
  2nd regional & 48 & 50 & 53 & 57 & 57 & 60 \\
  trust & 9 & 9 & 7 & 7 & 8 & 9 \\
  long-term & 2 & 2 & 2 & 3 & 3 & 3 \\
  others & 52 & 58 & 59 & 64 & 63 & 66 \\
  \hline
  total & 182 & 190 & 192 & 202 & 204 & 211 \\
  \hline
\end{tabular}
\caption{Numbers of city, regional, secondary regional, trust,
  long-term banks in each year.}
\label{tab:banks}
}
\end{table}

On the other hand, firms are all listed in the Japanese markets,
mostly consisting of large firms. Industrial sectors are classified
into 34 conventional sectors except banks and insurance, divided into
17 manufacturing and 17 non-manufacturing sectors. The numbers of
firms for each sector in each year are summarized in the table~\ref{tab:firms}.

\begin{table}[htbp]
\centering
{\small
\begin{tabular}{|l|r|r|r|r|r|r|}
  \hline
  sector-classification & 2005 & 2004 & 2003 & 2002 & 2001 & 2000 \\
  \hline
  Foods & 116 & 113 & 118 & 117 & 119 & 113 \\
  Textile Products & 52 & 55 & 58 & 61 & 58 & 63 \\
  Pulp \& Paper & 21 & 23 & 23 & 24 & 28 & 30 \\
  Chemicals & 151 & 154 & 158 & 160 & 162 & 159 \\
  Drugs & 21 & 23 & 25 & 29 & 32 & 36 \\
  Petroleum & 9 & 9 & 9 & 8 & 9 & 10 \\
  Rubber Products & 21 & 23 & 23 & 25 & 22 & 22 \\
  Stone, Clay \& Glass Products & 63 & 62 & 66 & 68 & 70 & 72 \\
  Iron \& Steel & 47 & 47 & 50 & 54 & 53 & 54 \\
  Non-ferrous Metal \& Metal Products & 99 & 100 & 107 & 110 & 112 & 121 \\
  Machinery & 191 & 196 & 200 & 205 & 207 & 204 \\
  Electric \& Electronic Equip. & 199 & 206 & 214 & 217 & 209 & 211 \\
  Shipbuilding \& Repair & 5 & 6 & 6 & 6 & 7 & 6 \\
  Motor Vehicles \& Auto Parts & 57 & 61 & 65 & 68 & 71 & 72 \\
  Transportation Equip. & 12 & 13 & 13 & 17 & 17 & 18 \\
  Precision Equip. & 45 & 47 & 45 & 46 & 45 & 43 \\
  Other Manufacturing & 97 & 95 & 93 & 98 & 94 & 89 \\
  \hline
  Fish \& Marine Products & 10 & 8 & 9 & 9 & 9 & 10 \\
  Mining & 6 & 7 & 9 & 8 & 9 & 9 \\
  Construction & 157 & 168 & 181 & 200 & 207 & 199 \\
  Wholesale Trade & 295 & 299 & 304 & 322 & 317 & 304 \\
  Retail Trade & 222 & 220 & 212 & 213 & 209 & 215 \\
  Securities houses & 19 & 13 & 15 & 15 & 14 & 12 \\
  Credit \& Leasing & 73 & 69 & 56 & 51 & 52 & 49 \\
  Real Estate & 100 & 93 & 84 & 79 & 71 & 57 \\
  Railroad Transportation & 28 & 31 & 31 & 32 & 32 & 34 \\
  Trucking & 35 & 35 & 32 & 31 & 28 & 30 \\
  Sea Transportation & 19 & 19 & 19 & 19 & 20 & 21 \\
  Air Transportation & 5 & 4 & 6 & 7 & 7 & 6 \\
  Warehousing \& Harbor Transportation & 39 & 38 & 36 & 35 & 37 & 34 \\
  Communication Services & 29 & 26 & 19 & 17 & 23 & 17 \\
  Utilities(Electric) & 9 & 9 & 8 & 8 & 8 & 7 \\
  Utilities(Gas) & 12 & 12 & 13 & 11 & 11 & 11 \\
  Services & 410 & 417 & 393 & 369 & 345 & 291 \\
  \hline
  total & 2,674 & 2,701 & 2,700 & 2,739 & 2,714 & 2,629 \\
  \hline
\end{tabular}
\caption{Numbers of firms in our datasets for each sector in each year
  (17 manufacturing and 17 non-manufacturing sectors).}
\label{tab:firms}
}
\end{table}

\newpage
\section{The network representation}\label{sec:nw}

\begin{figure}[htbp]
  \centering
  \includegraphics[width=0.7\textwidth]{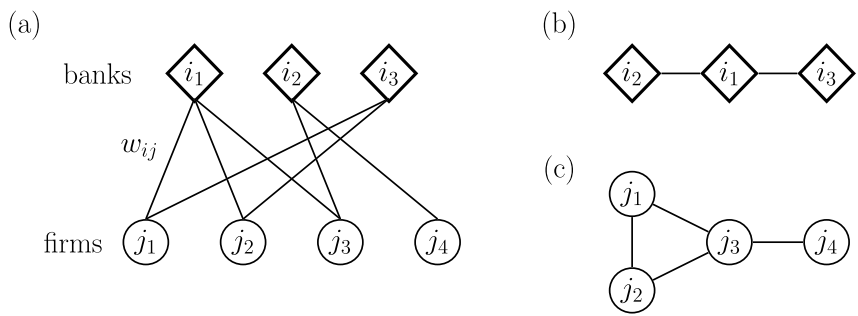}
  \caption{%
    (a)~Illustration of a bank-firm bipartite network.%
    (b)~Projected network on banks. (c)~Projected on firms.}
  \label{fig:bipartite}
\end{figure}

We represent the system as a network, by using an approach based on
the graph theory to analyze the structure of credit relationships in
the Japanese economic system. The network is defined as a set of nodes
and links and it is mathematically represented by a graph. In recent
years a large development of complex networks theory has been
observed. Many real systems have been represented as networks
\citep{caldarelli2007sfn, dorogovtsev2003enb}. Most of them show
scaling properties: they are scale-free networks, i.e. their
degree distribution is power-law tailed. In our case banks and firms
represent the nodes, while the links represent the credit
relationships between them. This type of networks is particular, being
composed by only two kinds of nodes, and is called {\it bipartite\/}
network. Figure~\ref{fig:bipartite}~(a) represents an example of the
bank-firm network.

Many empirical studies have been conducted in the field of bipartite
graphs (see e.g. \citet{peltomaki2006cbc,sneppen2004sms,guillaume2004bsa}). 
One can extract two networks from a bipartite network,
each one composed by just one kind of nodes: these two
networks are called {\it projected networks}, since they are obtained
as a projection of the initial graph in the subspace composed by nodes
of the same kind (see figure~\ref{fig:bipartite}~(b) and (c)).

A network is represented from a mathematical point
of view by an adjacency matrix. The element of the adjacency matrix
$a_{ij}$ indicates that a link exists between nodes $i$ and $j$; that
is, $a_{ij}=1$ if the bank $i$ provides a loan to the firm $j$;
otherwise $a_{ij}=0$. We can define a weighed adjacency matrix
$w_{ij}$ where $w_{ij}>0$ if the bank $i$ provides a loan to the firm
$j$ and the value of $w_{ij}$ is exactly the size of the loan;
otherwise $w_{ij}=0$.

The {\it degree\/} of a node $i$ is the number of its links and is
calculated by
\begin{equation}
  k_i=\sum_j a_{i,j}\ .
\end{equation}
The {\it neighbors\/} of a node $i$ is a set of nodes $j$ such that
$a_{ij}=1$, which is denoted by $\mathcal{V}(i)$.
The {\it strength\/} of a node $i$ is the total amounts of the weights
of its links and is calculated by
\begin{equation}
  s_i=\sum_j w_{i,j}\ .
\end{equation}

The {\it participation ratio\/} is a measure of the concentration of
the weight of a node versus its neighbors, and is defined by
\begin{equation}
  Y_i=\sum_j \left(\frac{w_{i,j}}{s_i}\right)^2\ .
\end{equation}
In the case of identical links (full homogeneity), the participation
ratio would be $Y_i=1/k_i$.
For a {\it main-bank\/} system, we expect the contracts of debt of
each firm to be concentrated, being the contract with the main
bank much more ``important'' with respect to those with the other
banks.

The {\it assortativity\/} is a measure of similarity among nodes and
it is defined as
\begin{equation}
  k_{nn}(i)=\frac{1}{k_i}\sum_{j \in \mathcal{V}(i)}  k_j\ .
\end{equation}

The {\it distance\/} $d_{ij}$ between two nodes $i,j$ is {\it the
  shortest} number of links to go from $i$ to $j$.  Therefore the
neighbors of a node $i$ are all the nodes $j$ which are connected
to that node by a single link ($d_{ij}=1$). Using the adjacency
matrix this can be written as
\begin{equation}
  d_{ij}=\textrm{min}\left[\sum_{k,l \in \mathcal{P}_{ij}} a_{kl}\right]
  \ ,
\end{equation} 
where $\mathcal{P}_{ij}$ is a path connecting node $i$ and node
$j$. The {\it diameter\/} of a graph is given by the maximum of all
distances between pairs.


In the following we apply these tools to our dataset. It is mentioned
that \citet{fujiwara2009stc} examined a similar dataset to quantify
the dependency and influence between banks and firms by using the
weights defined above and assuming a diffusion process in the
bipartite graph. The present paper focuses on statistical properties
of credit topology and weights, and further on extraction of cohesive
groups or communities in the network, and also an extensive analysis
of hierarchical structures.

\section{The Banks-Firms credit network}\label{sec:nw_bf}

The {\it average degree\/} of firms is $\left\langle
  k_f\right\rangle=8$, while the {\it average degree\/} of banks is
$\left\langle k_b\right\rangle=120$; the average strength of firms
$\left\langle s_f\right\rangle=2.15\times 10^4$ million yen, while
that of banks is $\left\langle s_b\right\rangle=3.6\times 10^5$
million yen.

\begin{figure}[htbp]
  \centering
  \includegraphics[width=0.45\textwidth]{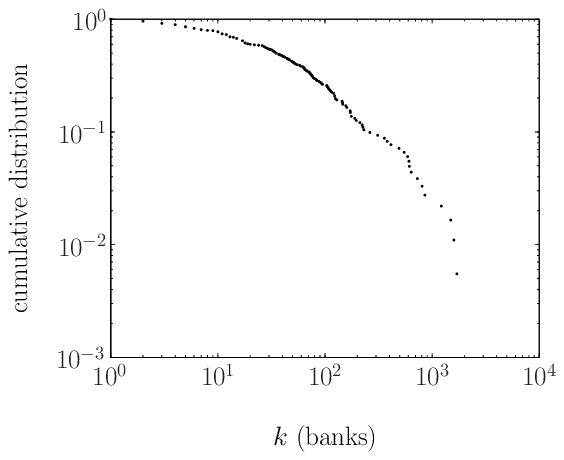}\qquad%
  \includegraphics[width=0.45\textwidth]{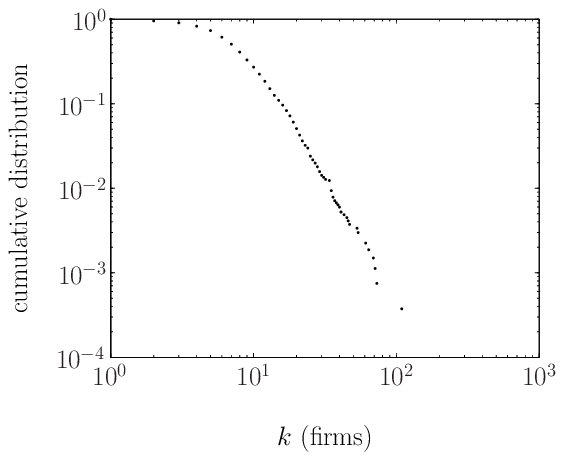}
  \caption{%
    Degree cumulative distribution of banks (left panel) and firms
    (right panel). The cumulative distribution has a power-law tail,
    $P^{>}(k) \propto k^{-\mu}$ in both
    cases. For banks $\mu=0.9 \pm 0.1$, for firms the estimated
    parameter is $\mu=2.6 \pm 0.1$. The estimation of the exponent is
    done by maximum likelihood method (Hill's estimate) here and
    hereafter.}
  \label{fig:Pdk}
\end{figure} 

\begin{figure}[htbp]
  \centering
  \includegraphics[width=0.45\textwidth]{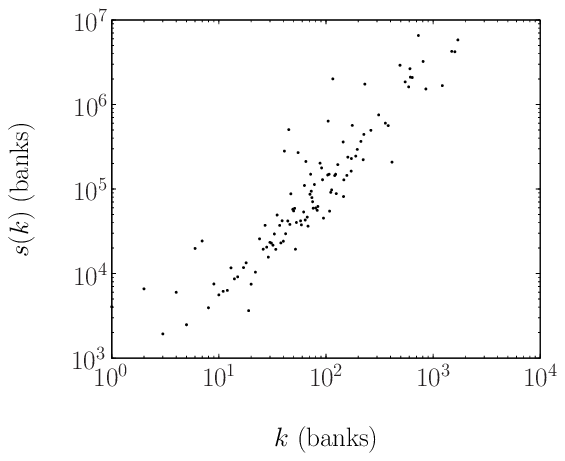}\qquad%
  \includegraphics[width=0.45\textwidth]{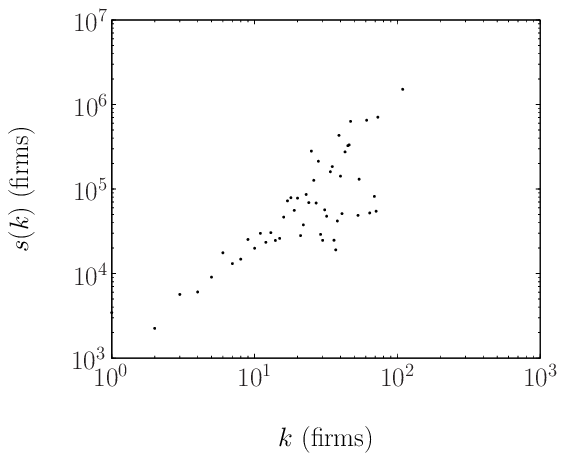}\\
  \includegraphics[width=0.45\textwidth]{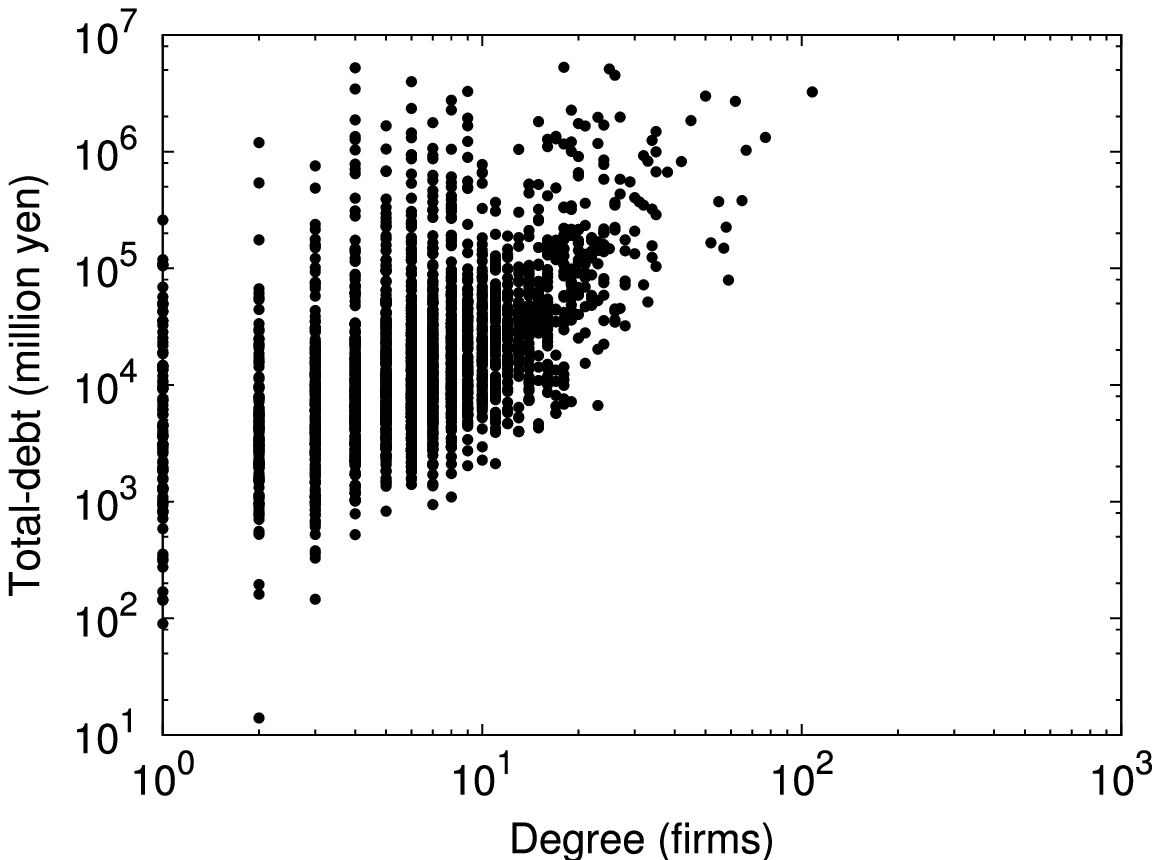}
  \caption{%
    Top panels: strength $s$ vs.~degree $k$ of banks (left panel) and
    firms (right panel), considering total contracts. Bottom panel:
    scatter plot for degree (firms) and total-debt.}
  \label{fig:sdk}
\end{figure} 

Figure~\ref{fig:Pdk} shows the distribution of the degree for banks
(left panel) and firms (right panel). The maximum degree of the banks
is 1,706, and that of the firms is 109, with very heterogeneous behavior among
banks and among firms. In particular, after having conditioned it for
the firm's size, we found that many firms prefer single lending
whereas others adopt a multiple lending strategy.

In figure~\ref{fig:sdk} (top panels), we show the scaling of the
strength versus the degree. In the case of banks the linear
correlation coefficient between $s$ and $k$ is 0.8, while that for firms
is 0.4. This signals the presence of a weak
link between the amount of credit firms demand the banks for and the
number of banks they ask to. The firms with large amount of
borrowings prefer multiple links (in agreement with
\citeauthor{ogawa2007jfp},~\citeyear{ogawa2007jfp}), but multi-lending
is present also among firms with a lower amount of borrowings. In
figure~\ref{fig:sdk} (bottom panel) we observe the scatter plot for
the degree of firms versus their total-debt. By calculating a rank
correlation (Kendall's $\tau$), we found that $\tau=0.382$
($27.0\sigma$) (where $\sigma$ is what is expected from the null
hypothesis that there is no association between the rank of degree
and that of total-debt), which implies
significantly positive correlation.

\begin{figure}[htbp]
  \centering
  \includegraphics[width=0.45\textwidth]{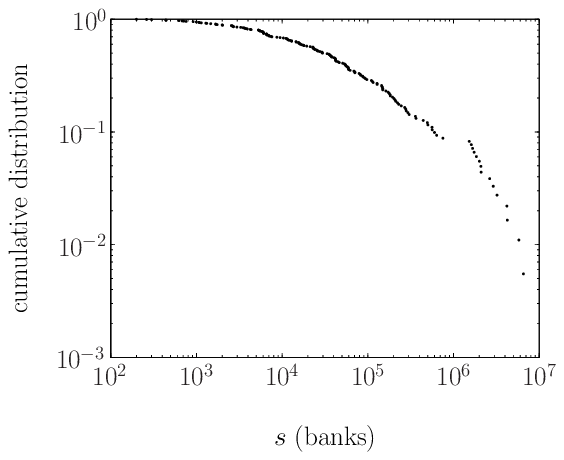}\qquad%
  \includegraphics[width=0.45\textwidth]{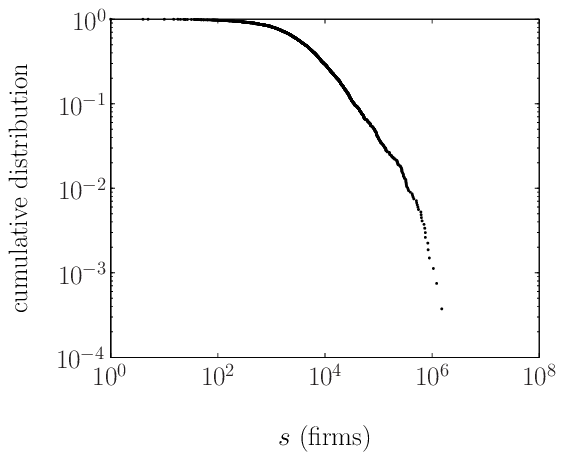}
  \caption{%
    Strength cumulative distribution of banks (left) and firms
    (right), considering total contracts. The best fit, in the
    intermediate range of values, is $P^{>}(s) \propto s^{-\mu}$ in
    both cases. For firms the estimated parameter is $\mu=0.86 \pm
    0.03$, for banks $\mu=0.51 \pm 0.05$.}
  \label{fig:Pds}
\end{figure}

In figure~\ref{fig:Pds}, the distribution of the strength for total
contracts is plotted. The maximum strength of banks is $6.5\times
10^6$ million yen and $1.5\times 10^6$ for firms, and there is no
difference in the plots when long-term and short-term loans are
considered.

\begin{figure}[htbp]
  \centering
  \includegraphics[width=0.45\textwidth]{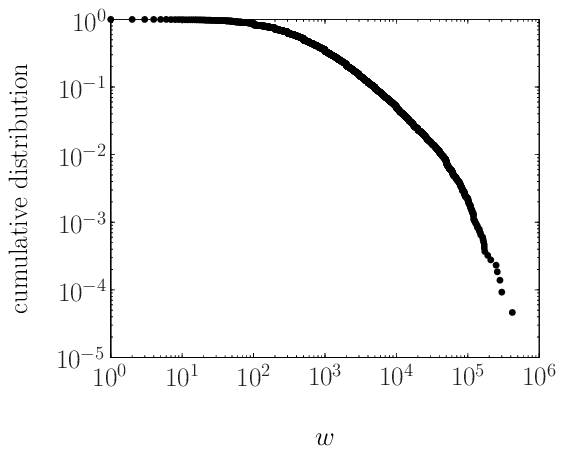}\qquad%
  \includegraphics[width=0.45\textwidth]{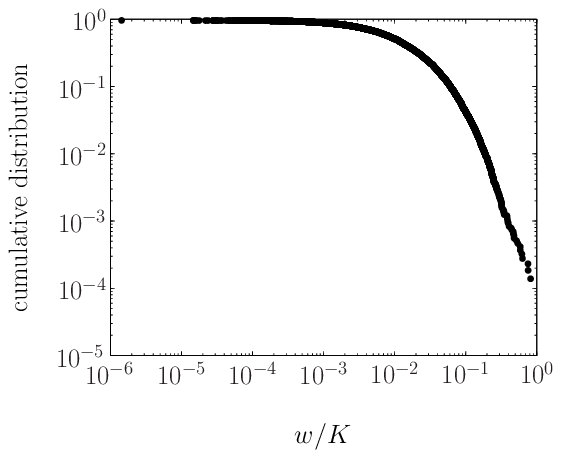}
  \caption{%
    Comparison of weights distributions (left panel) with the
    distribution of weights renormalized by the capital size (right
    panel). The best fit is $P^{>}(w)\propto w^{-\mu}$. In the left
    plot estimated parameter is $\mu=0.95 \pm 0.01$, in the right one
    $\mu=2.38 \pm 0.08$.}
  \label{fig:Pdw}
\end{figure}

Once one looks into the differentiation of the lending contracts for
all the firms, a fat tail distribution of the amount of the contracts
emerges (figure~\ref{fig:Pdw}, left panel). This heterogeneity is not
a consequence of the heterogeneity of firms' sizes. In fact,
after normalization, we can still observe a fat-tailed distribution
in the right panel of figure~\ref{fig:Pdw}.

\begin{figure}[htbp]
  \centering
  \includegraphics[width=0.50\textwidth]{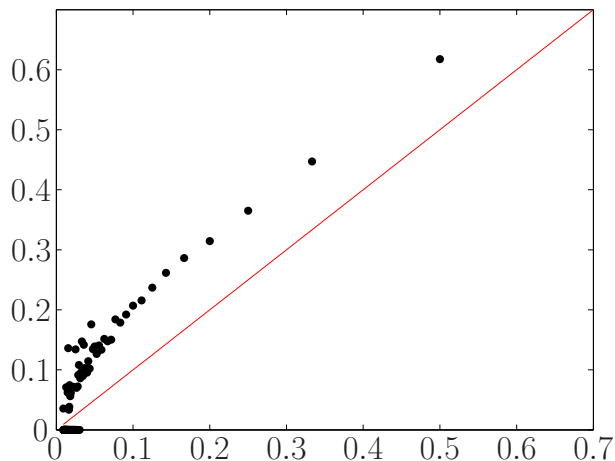}
  \caption{%
    Participation ratio $Y$ of firms vs.~$1/k$.}
  \label{fig:Y}
\end{figure} 

Which is the differentiation inside the set of contracts of firm by
firm? The {\it participation ratio\/} is a measure of the
heterogeneity of the amount of debt of a certain firm versus its
creditors, i.e. if the sizes of different loans are roughly of a
similar size or not. The ratio for a firm $f$ is
\begin{equation}
  Y_f=\sum_b \left(\frac{w_{f,b}}{s_f}\right)^2\ .
\end{equation}
In figure~\ref{fig:Y}, the actual participation ratio is represented
with black dots, while the red line represents the ``homogeneous''
case $Y_f=1/k_f$. As expected, the dots do not overlap with the
homogeneous line.  Let us note that heterogeneity is stronger in the
case of weak multiple lending (low $k$, right side of the $x$ axis).

No striking differences in the distributions of firms-degree are
observed when one considers separately long- and short-term contracts:
the average degree $k$ is 6.7 (for the long-term loan) and 5.7
(short-term), while the maximum $k$ is 108 (long-term) and 54
(short-term). The linear correlation coefficient between the degree
and the total amount of borrowings for firms is 0.19 (short term) and
0.39 (long term), while they increase quite a lot for banks to 0.79
(short term) and 0.88 (long term) respectively. It looks as if there
is (no) statistically robust link between long- (short-) term
contracts and multi-lending, because limited information induces risk
diversification.

\begin{figure}[htbp]
  \centering
  \includegraphics[width=0.45\textwidth]{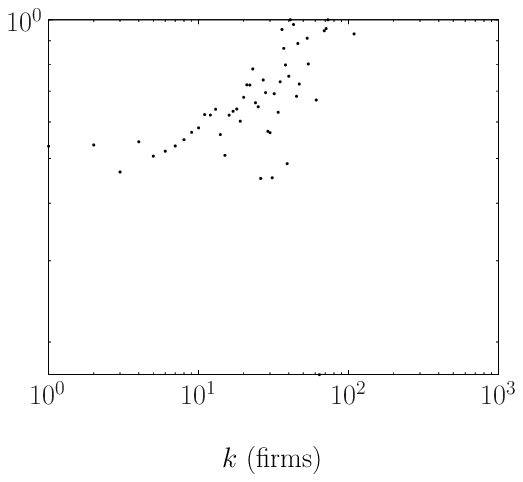}\qquad%
  \includegraphics[width=0.45\textwidth]{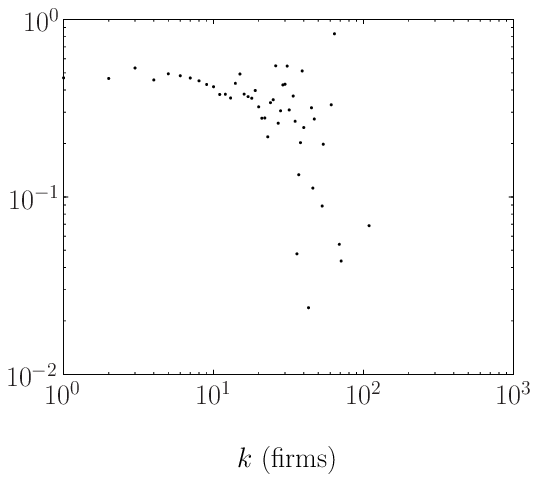}
  \caption{%
    Relative long (left) and short (right) term strength vs. degree of
    firms. Firms with high degree subscribe long-term contracts.}
  \label{fig:longvsshort}
\end{figure} 

If one analyzes the percentage of short-term or long-term contracts
with respect to the total amount of borrowings versus degree $k$ (as
plotted in figure~\ref{fig:longvsshort}), a decreasing tendency of the
ratio short/long can be emphasized.

\section{Hierarchical clustering of co-financing banks}\label{sec:nw_banks}

From the network of banks and firms, we can extract the network of the
{\it co-financing\/} banks with the method of projected network
\citep{masi2007bft}. The obtained bank network is defined as a
weighted network, only populated by banks, in which two banks are
linked if they finance the same firm; therefore, the weight $w$ of the
link is the number of firms they both finance. The banks are divided
in 6 subgroups depending on the kind of bank, as shown in table~\ref{tab:classbanks}.

\begin{table}[htbp]
\centering
{\small
\begin{tabular}{|l|c|r|}
  \hline
  large sector & color & description \\ \hline
  1 & black & long-term credit bank \\ \hline
  2 & blue & city banks\\ \hline
  3 & green & regional banks \\ \hline
  4 & yellow & trust banks\\ \hline
  5 & orange & secondary regional banks\\ \hline
  6 & white & the rest of banks\\ \hline
\hline
\end{tabular}
\caption{%
  Classification of banks}
\label{tab:classbanks}
}
\end{table}

Instead of considering the whole weighted adjacency matrix $W$ and the
whole network, we analyze a tree with only $N-1$ links, which select
the most of important links of the matrix $w_{i,j}$. The algorithm
used to construct the tree of banks is the Minimal Spanning Tree (MST)
(e.g. \citet{cormen2001ia}) (see %
\citet{mantegna1999hsf} for an early
application in financial market). We consider a set of $N$ banks and the weighted
matrix $w_{i,j}$ of the number of contracts in common among them. We
define a distance between a pair of banks
\begin{equation}
  d_{i,j}=1-\tilde{w}_{i,j}\ ,
\end{equation}
where $\tilde{w}_{i,j}=w_{i,j}/w_{\scriptsize\textrm{max}}$, and
$w_{\scriptsize\textrm{max}}$ is the maximum among all the weights
of the links. Then the MST is calculated in the following way:
\begin{itemize}
  \item rank by increasing order the $N(N-1)/2$ values of $d_{i,j}$
  \item pick the pair corresponding to the smallest $d_{i,j}$ and
    create a link between these two banks
  \item pick the pair corresponding to the second higher $d_{i,j}$ and
    create a link between these two banks
  \item repeat the operation {\em unless} adding a link between the
    pair under consideration creates a cycle, in which case skip that
    value of $d_{i,j}$.
\end{itemize} 
In this way we find a tree containing the strongest links of the
original weighted matrix $w_{i,j}$.

\begin{figure}[htbp]
  \centering
  \includegraphics[width=0.80\textwidth]{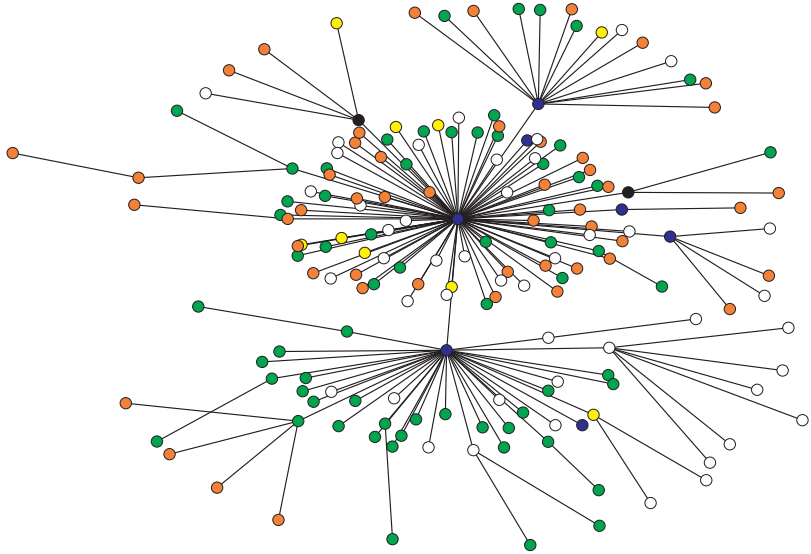}
  \caption{%
    Minimal Spanning tree for year 2004 with 178
    banks: the colors indicate the kind of bank.}
  \label{fig:MST}
\end{figure}

In figure~\ref{fig:MST}, we plot the MST. This is the backbone of
co-financing relationships in Japan. In the year 2004, the hubs are
Sumitomo Mitsui Banking Corporation (center), the Bank of
Tokyo-Mitsubishi, Ltd (bottom) and UFJ Bank, Ltd. (top). The three
hubs structure allows separating the bank system in three sub-graphs:
the failure of one of the three largest banks can cause a huge impact
in each corresponding subsystem that is divided by the clusters.
We observe clusters with strong geographical characterization: the
nodes' neighbors in the tree are in the same geographical region. We
have clear hierarchical clustering, where the hubs are Tokyo
Mitsubishi, Sumitomo Mitsui and UFJ bank. These three hubs,
Tokyo-Mitsubishi, UFJ and Sumitomo Mitsui, are the largest banks in
Japan until the year 2006, when the former two banks were merged into
a single and largest bank. We may recognize branches from the Chubu
region, pairs of banks from Tohoku, Chubu, Kantou, Kyushu, triads from
Chugoku, Okinawa; also some pairs of institutes of life insurance.

We can observe a very clear geographical location of the clusters. We
may define the color of the nodes considering their region (table~\ref{tab:geobanks}).

\begin{table}[htbp]
\centering
{\small
\begin{tabular}{|r|r|r|}
  \hline
  group & color & region \\
  \hline
  0 & white & not regional \\ \hline
  1 & black & Hokkaido and Tohoku \\ \hline
  2 & blue & Kantou \\ \hline
  3 & green & Chubu \\ \hline
  4 & yellow & Kinki \\ \hline
  5 & orange & Chugoku \\ \hline
  6 & red & Shikoku \\ \hline
  7 & brown & Kyushu\\ \hline
  \hline
\end{tabular}
\caption{%
  Geographical classification of banks.}
\label{tab:geobanks}
}
\end{table}

\begin{figure}[htbp]
  \centering
  \includegraphics[width=0.80\textwidth]{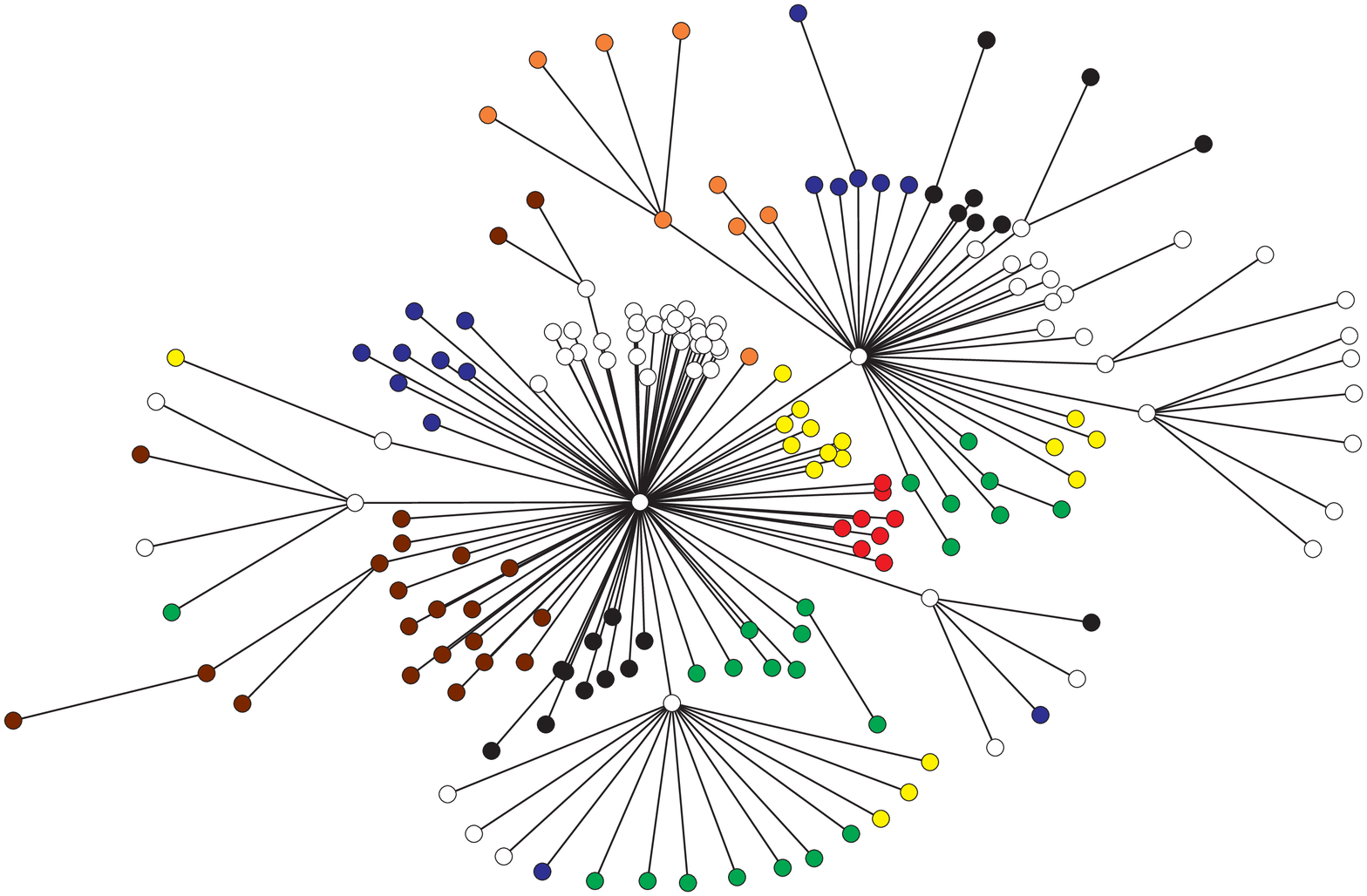}
  \caption{%
    Minimal Spanning tree for year 2004 with 178 banks: the colors
    indicate the geographical locations.}
  \label{fig:MSTgeo}
\end{figure}

The MST obtained with this definition of colors is represented in
figure~\ref{fig:MSTgeo}.

A look at the MST reveals that several of its portions can be well
understood. We can interpret the tree by making some considerations:
\begin{itemize}
\item The main reason why banks and financial institutions have
  borrowers in common is that they do the lending activity in the same
  geographical regions. In particular, very frequently, regional banks
  have common sets of borrowers in the same regional places. This is
  because Japanese firms have customarily borrowed from more than one
  bank. This is in contrast with the US, where many small and
  medium-size firms have single borrowings. Moreover, traditionally
  major banks operate mainly in the urban regions of Tokyo, Osaka and
  Nagoya, so their common borrowers are quite similar. On the other
  hand, in recent years, major banks have increased the number of
  branch offices in suburban areas, so they have begun to share common
  borrowers with regional banks.
\item The set of common borrowers is also explained considering that
  in Japan the lending activity is based on {\it keiretsu\/}, between
  major and regional banks, between banks and trust-banks, and between
  banks and insurance companies (life-, fire- and marine-insurances).
  Frequently, big firms in industrial business conglomerates (groups
  partly remnant from the pre-war {\it zaibatsu\/}) have borrowed
  from closely related conglomerates of financial businesses of banks,
  trust-banks and insurance companies.
\item Two other reasons may determine what has been observed. The
  first one is that when foreign-affiliated (-owned) insurance
  companies lend to individuals (these activities are extended to
  lending to firms as well), they are supported by branch offices of
  regional banks. The second one is that {\it keiretsu\/} exists
  between major banks and regional banks. In the same {\it
    keiretsu\/}, human resources, financial technologies and operating
  systems are shared. This can possibly yield opportunities for
  sharing common borrowers. In the MST different main branches of the
  tree correspond to the {\it keiretsu\/} between major and regional banks.
\item Banks had been customarily owners of firms' equities in
  correlation with shares of lending to those firms. In other words,
  lending relationships are associated with particular ownership
  relations. This activity would result in common lending followed by
  common shareholding.
\item Finally, firms happen to be in the same {\it syndicated loan}.
  This is a large loan in which a group of banks work together to
  provide funds for a borrower. There is usually one leading bank,
  which is called {\it arranger\/} and is often a major bank, which
  takes a percentage of the loan and syndicates the rest to other
  banks.
\end{itemize}

\section{Co-financed firms network}\label{sec:nw_firms}

We can project the bipartite network in the subspace of firms,
obtaining the co-financed firms' network. This strongly connected
network was created in 2004 by 2,661 firms, linked to each other by
almost 3 millions (2,881,763) links (as the number of possible links
is 3,539,130 this is more than 80\% of all existing possible links
among firms). The average connectivity is $k=2,164$. If the lending was
sectorial (i.e. certain banks finance certain firms (for example firms
of the same industrial sector), while other banks finance other firms,
this would imply, in the projected space, the formation of grouping of
cofinanced firms (properly said communities in the network
literature). On the contrary, in the empirical case, we observe that
each firm is connected to many other ones and that there are not
communities. This is a sign of the fact that the lending is not
sectorial.

We aggregate several sectors of table~\ref{tab:firms} into six groups: 1) Foods,
Chemicals, Drugs; 2) Iron, Steel, Non-ferrous Metals, Metal Products;
3) Motor Vehicles, Auto Parts, Transportation Equip., Shipbuilding,
Repair; 4) Machinery, Electric and Electronic Equip., Precision
Equip., other Manufacturing; 5) the rest of manufacturing sectors, and
6) all of non-manufacturing sectors.

\begin{figure}[htbp]
  \centering
  \includegraphics[width=0.40\textwidth]{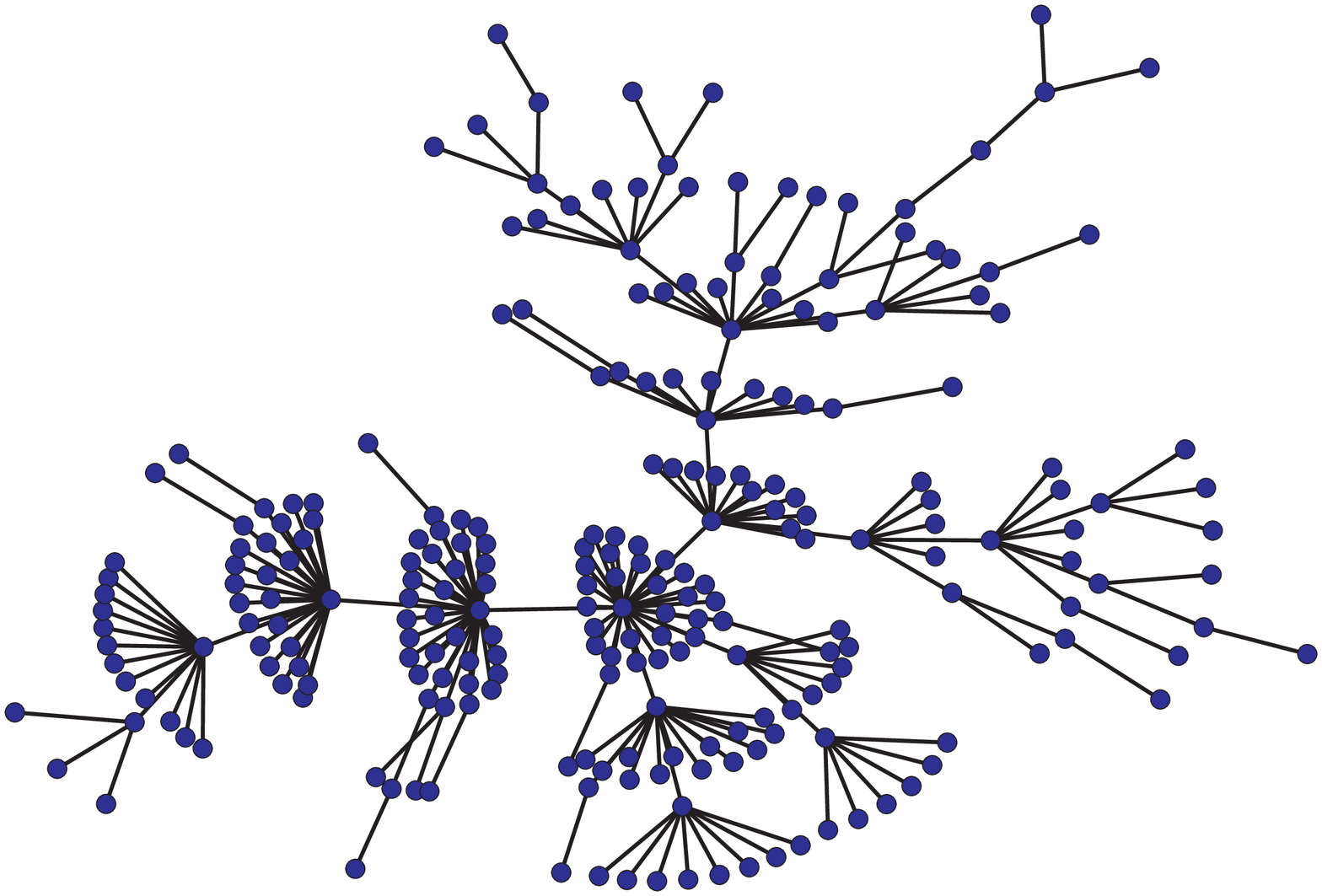}\quad%
  \includegraphics[width=0.40\textwidth]{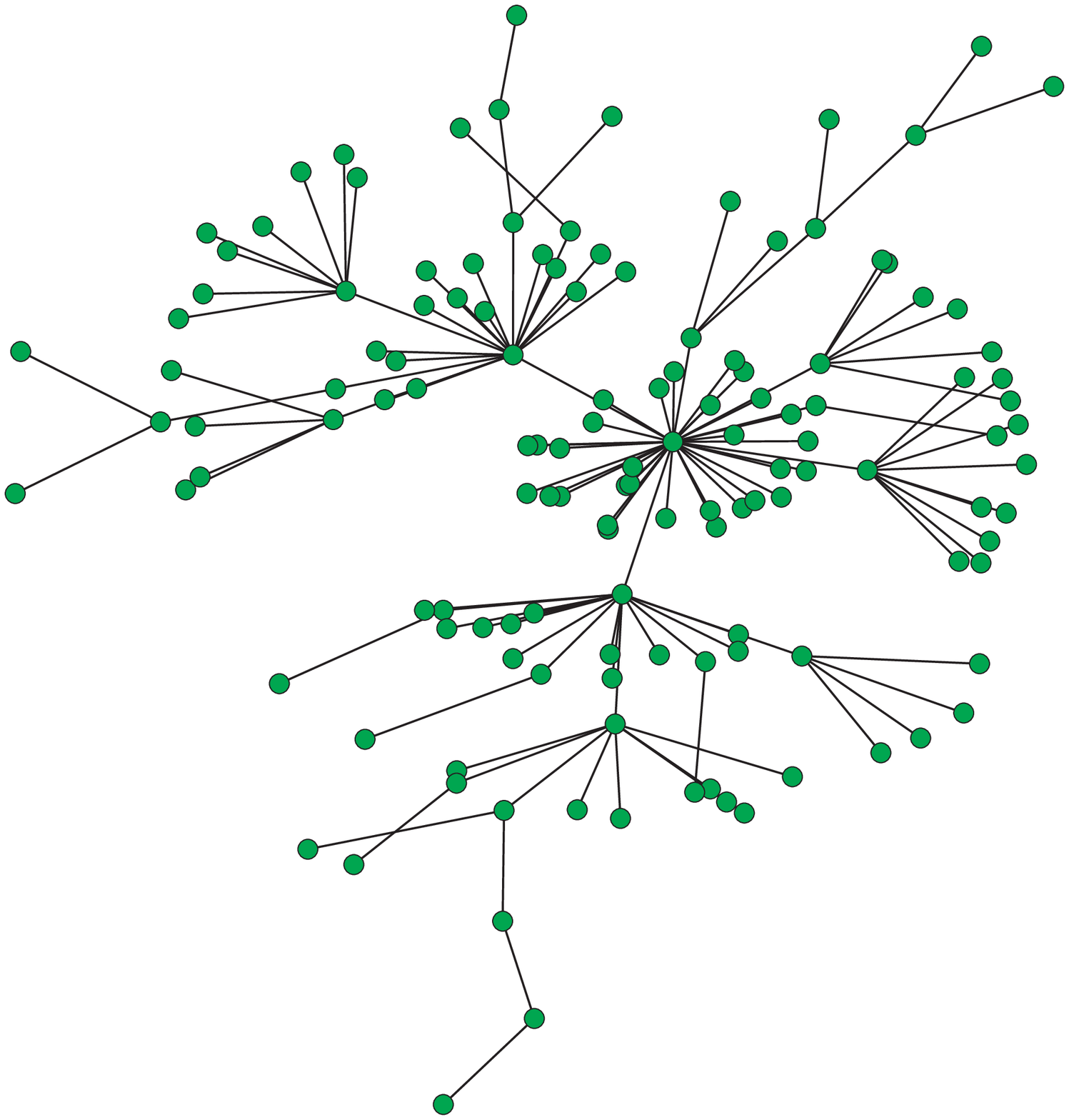}\\
  \includegraphics[width=0.40\textwidth]{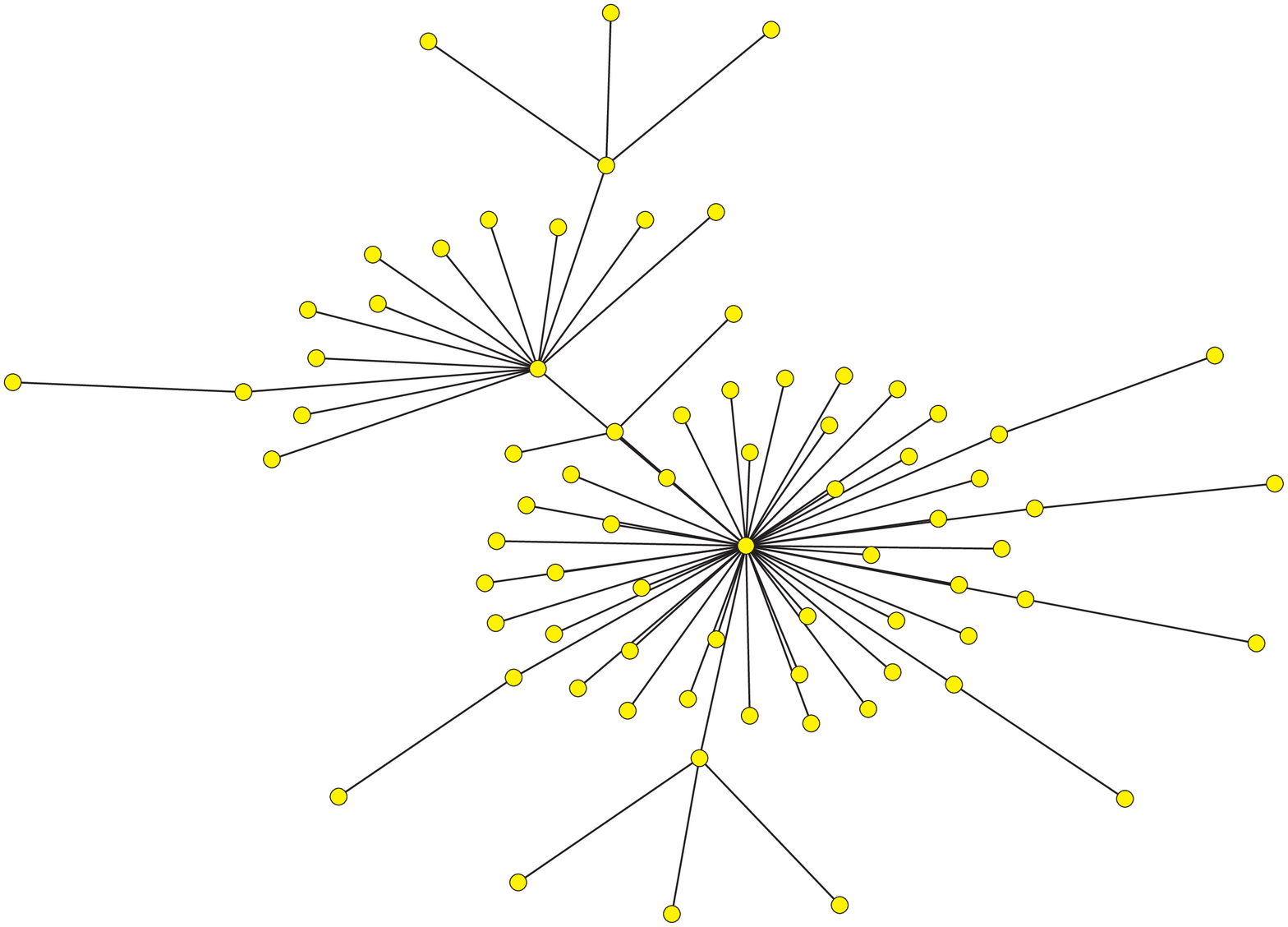}\quad%
  \includegraphics[width=0.40\textwidth]{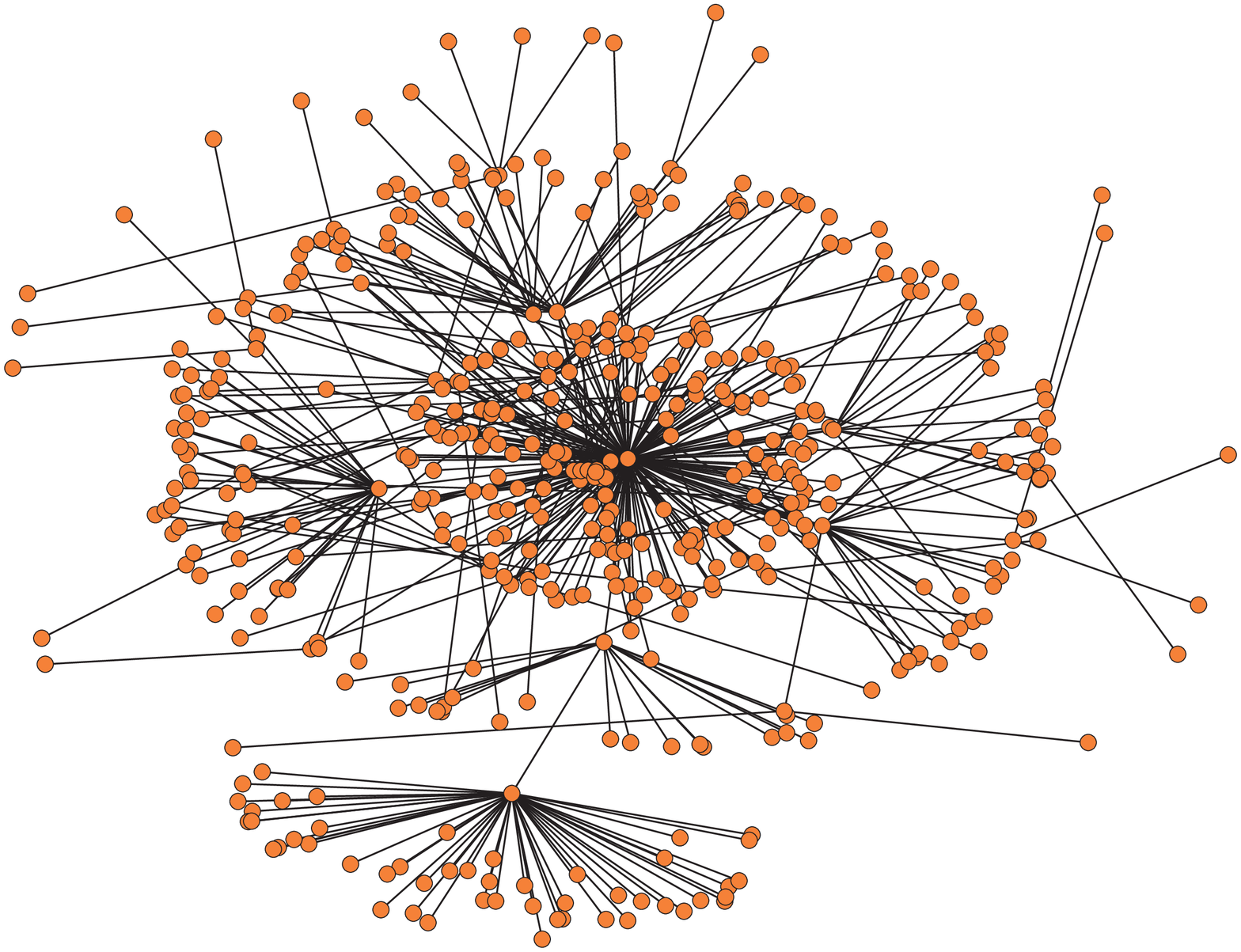}\\
  \includegraphics[width=0.40\textwidth]{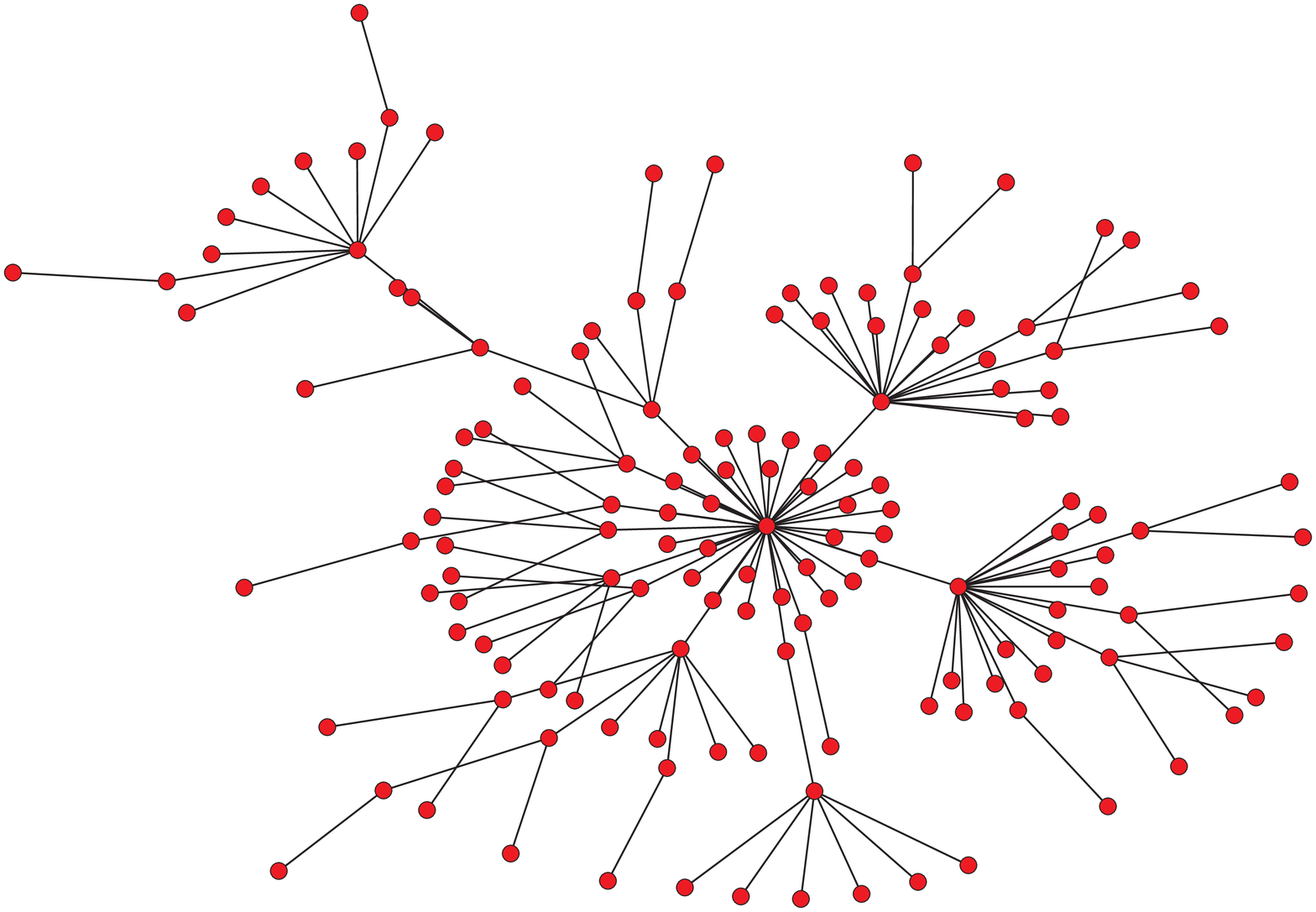}\quad%
  \includegraphics[width=0.40\textwidth]{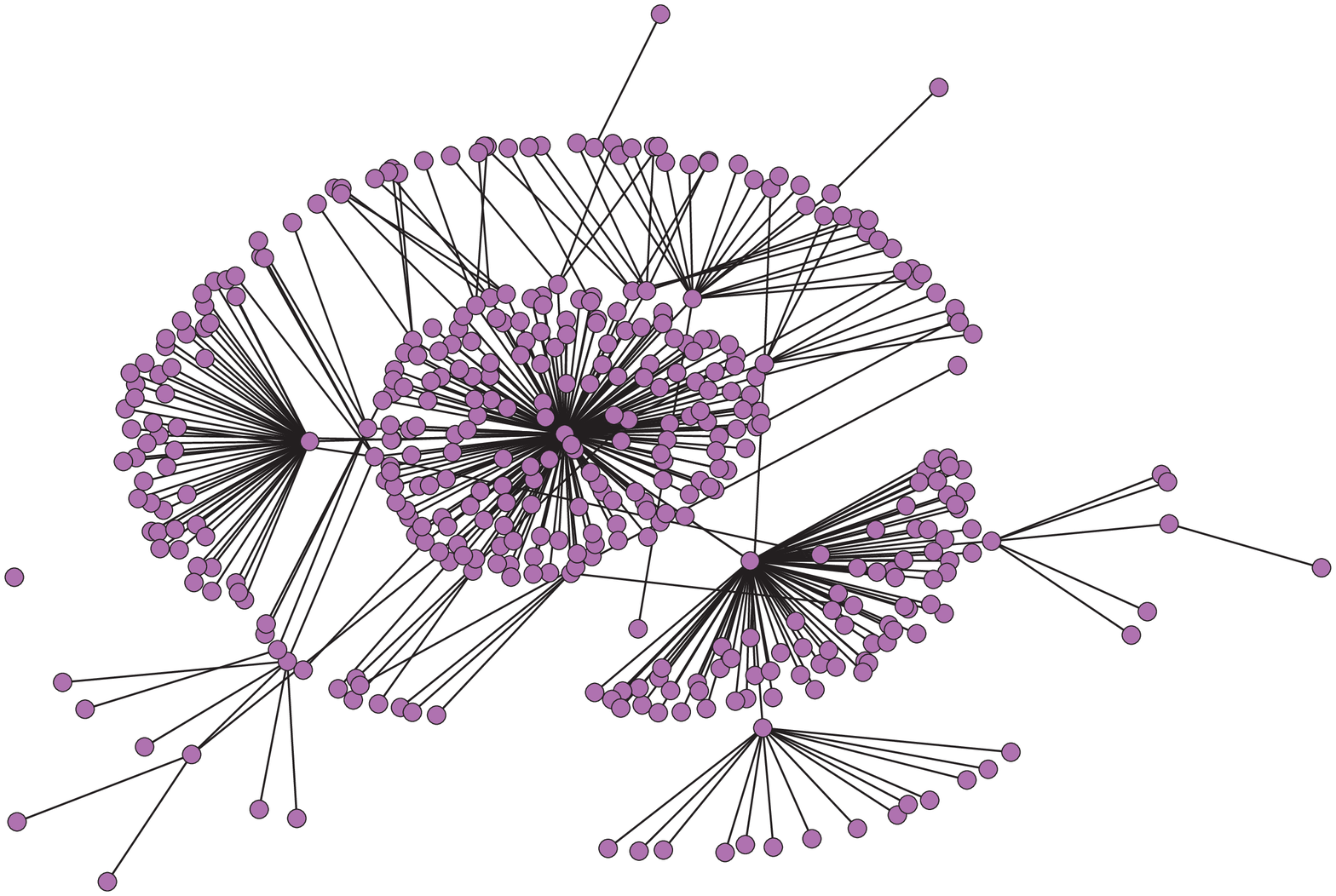}
  \caption{%
    MST of firms sector by sector.}
  \label{fig:MSTfirms}
\end{figure}

In the following we extract the sub-networks of firms belonging to the
same group. We calculated the MST for each sector. The trees
obtained are reported in figure~\ref{fig:MSTfirms}. For each tree, the
hubs are indicated in the caption of the figure.

Investigating the clusters, one may observe a regular pattern, except
for the first group of sectors (Foods, Chemical and Drugs, panel top
left): there exists a very big firm as a hub, which has many connected
firms and from 2 to 5 large firms. In turn, these firms create
sub-hubs, i.e. they are connected to other smaller firms constituting
an autonomous tree. Note that asymmetric information may be the cause
of such a configuration. On the one hand, in fact, banks tend to
specialize in providing credit to some sectors only (they diversify
the risk by investing in different industrial sectors); but also there
is a geographical specialization, which leads to the birth of the
sub-hubs.

\section{Financial status and topology}\label{sec:financial_status}

This section investigates the effect of the financial status of the
firms on the topology structure as well as of banks. Data show that
debt, asset and {\it DAR\/} (debt on asset ratio) are correlated with
the total degree of the firm, although not very strongly. This signals
the presence of strong heterogeneity among firms with similar degrees.

In order to test the statistical significance of the correlation
values, we compute the probability $p$ of obtaining a correlation as
large as the observed value by random chance, when the true
correlation is zero. If $p$ is small, say less than 0.05, then the
correlation is significant. The $p$-value is computed by transforming
the correlation to create a $t$-statistic having $N-2$ degrees of
freedom, where $N$ is the number of rows of $X$. In the following
table~\ref{tab:corr_status} we observe the correlation values among degree, debt, asset and
DOA.

\begin{table}[htbp]
\centering
{\small
\begin{tabular}{|cc|r|r|}
  \hline
  variable 1 & variable 2 & correlation & $p$  \\
  \hline
  debt & degree & 0.32 & $<0.001$ \\
  asset & degree & 0.29 & $<0.001$ \\
  DAR & degree & 0.24 & $<0.001$ \\
  debt & asset & 0.97 & $<0.001$ \\
  \hline
\end{tabular}
\caption{%
  Correlation among degree, debt, asset and DOA.}
\label{tab:corr_status}
}
\end{table}

\begin{figure}[htbp]
  \centering
  \includegraphics[width=0.45\textwidth]{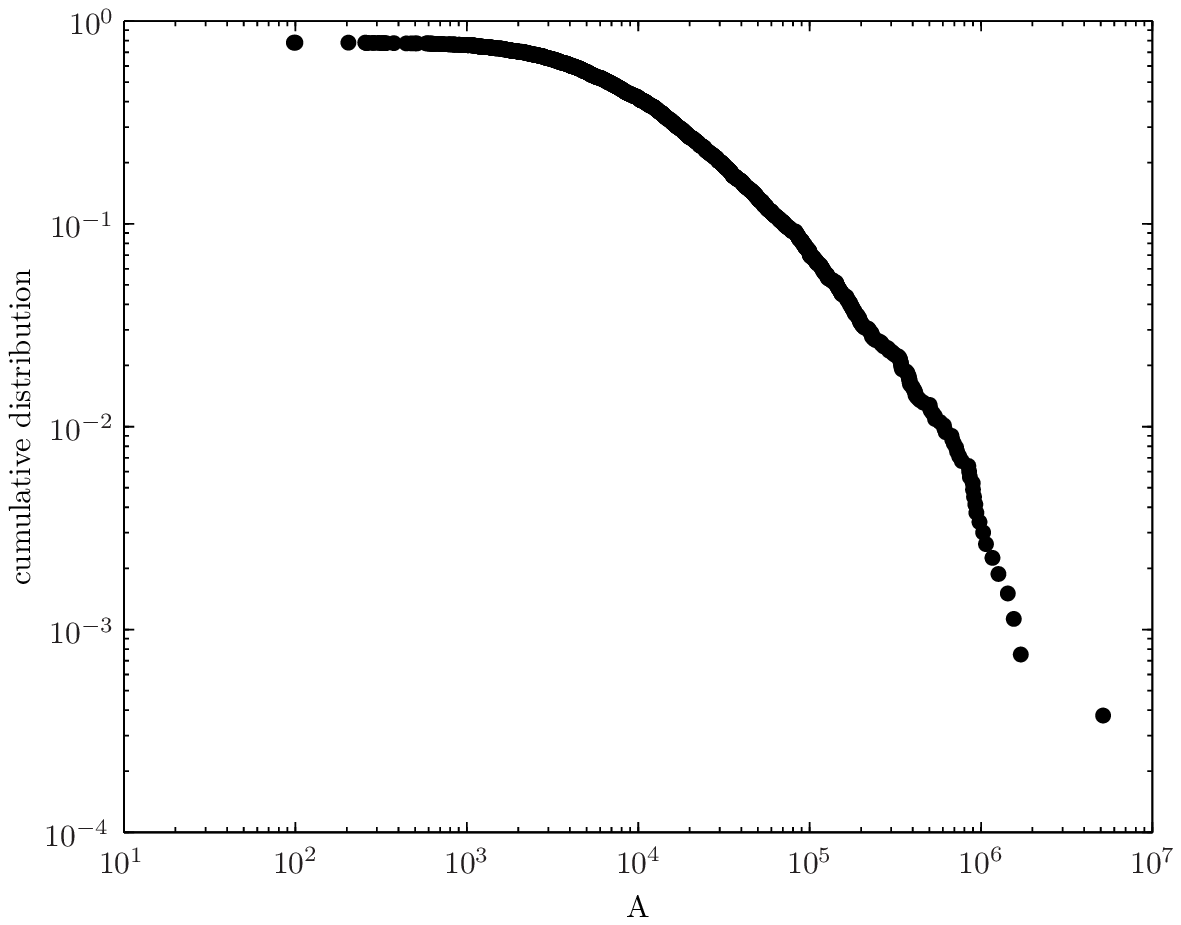}\qquad%
  \includegraphics[width=0.45\textwidth]{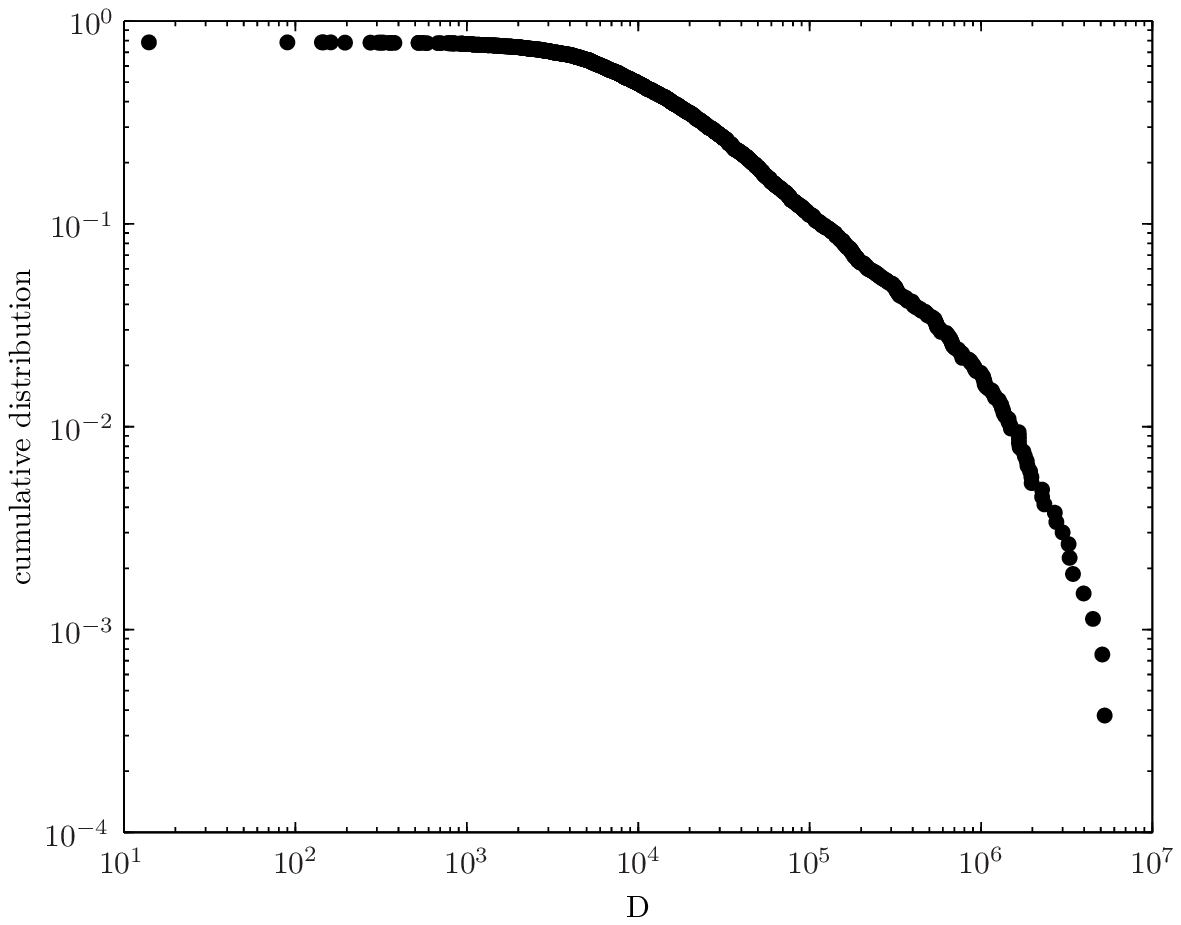}
  \caption{%
    Asset $A=K-D$ and debt $D$ cumulative distributions for the year
    2004. The best fit is $P^{>}(x) \propto x^{-\mu}$ in both cases.
    In the left plot the estimated parameter is $\mu=0.82 \pm 0.03$,
    in the right one $\mu=0.74 \pm 0.02$.}
  \label{fig:PdA}
\end{figure}

\begin{figure}[htbp]
  \centering
  \includegraphics[width=0.50\textwidth]{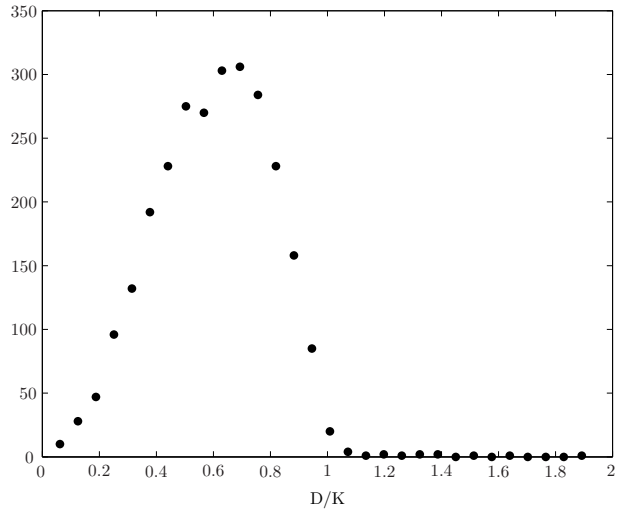}
  \caption{%
    Debt on capital distribution for the year 2004. The distribution
    of liabilities is power law distributed as observed by 
    \citep{fujiwara2004zlf}.}
  \label{fig:PdDAR}
\end{figure}

The debt and the size (asset) of the firm are highly correlated, as
pointed out in a previous work by \citet{fujiwara2004zlf}. The
distributions of asset, debt and debt on capital of our sample are
plotted in the figure~\ref{fig:PdA} and figure~\ref{fig:PdDAR}. The
firms are divided in 5 classes based on the different asset value. The
classes are built with the aim to obtain 5 equally populated classes.

\begin{figure}[htbp]
  \centering
  \includegraphics[width=0.50\textwidth]{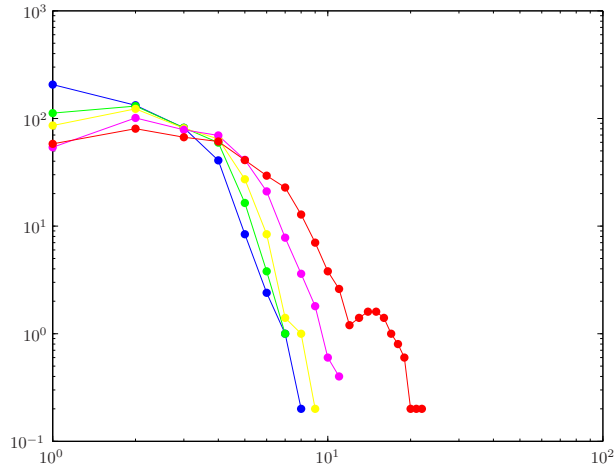}
  \caption{%
    $P(k)$ distribution for the year 2004: blue dot are the smallest
    firms and, increasing their asset value, we define the colors
    green, yellow, magenta, red.}
  \label{fig:Pdkcol}
\end{figure} 

\begin{figure}[htbp]
  \centering
  \includegraphics[width=0.50\textwidth]{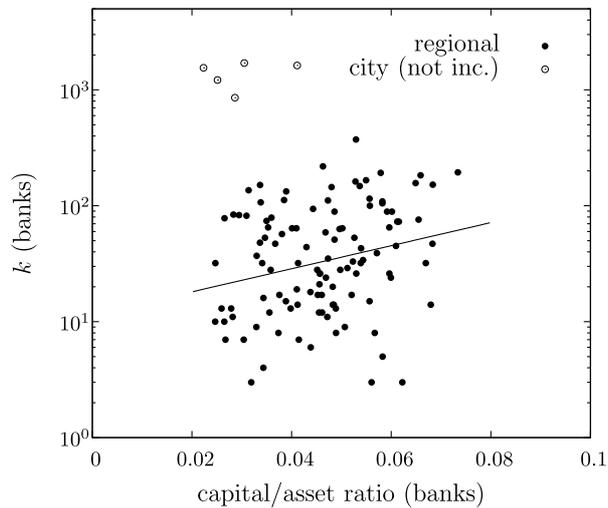}
  \caption{%
    Scatter plot for the capital-to-asset rations and the degrees of
    banks. Filled dots are the regional and 2nd regional banks, while
    circles are city banks. The linear regression for the ratio and the
    log of degree is shown by a solid line, where the city banks are
    not included in the regression analysis.}
  \label{fig:bank_deg-c-a}
\end{figure} 

In figure~\ref{fig:Pdkcol}, we plot the degree distributions for firms
belonging to each of the 5 classes. Firms' behavior of the same
classes is very heterogeneous: the number of contracts in each class
is variable, even if the value of $P(k)$ shifts toward higher values
of $k$, when the size of the firms increase: multi-lending is present
in both small and large firms, even if the bigger firms have a larger
number of creditors.

In addition, we checked the widely recognized hypothesis that the
balance-sheet conditions of financial institutions also affect the
bank-firm relationships. We examined the relation between the
capital-to-asset ratio of banks and their degrees.
Figure~\ref{fig:bank_deg-c-a} shows the scatter plot of the
capital-to-asset ratios and degrees for the regional banks (filled
dots) and also for the city banks (circles). The latter group
obviously has a different distribution, so we focused only on the
regional banks and confirmed that there exists a weakly positive
correlation, $R=23.4\%$ ($p$-value $<10^{-2}$).

\section{Conclusive Remarks}\label{sec:conclusion}

In this conclusive section, we would like to emphasize how a {\it new
  tool\/} for economic policy emerges from the network analysis,
namely the issues of {\it stabilization\/} of the financial system by
preventing a {\it financial crises}, with its {\it propagation\/} and
{\it amplification\/}, or {\it domino effects}. Real economies are
composed by millions of interacting agents, whose distribution is far
from being stochastic or normal. The Japanese credit market shows that
several hubs exist, i.e. banks and firms with many connections: the
distribution of the degree of connectivity is {\it scale-free}, i.e.
there are a lot of firms with 1 or 2 links, and very few firms with a
lot of connections well described by a scale-free distribution. Let us
assume the Central Authority has to prevent a {\it financial
  collapse\/} of the system, or the spreading of it (the so-called
{\it domino effect\/}). Rather than looking at the ``average'' risk of
bankruptcy, and to infer it would represent the stability of the
system, the network analysis of the real system tells us to
investigate the different sub-systems of the global economy and to
intervene to prevent failures and their spread. Instead of a
helicopter drop of liquidity, one can make ``targeted'' interventions
to a given agent or sector of activity: \citet{fujiwara2008cfb} shows
how to calculate the probability of going bankrupt by {\it solo\/},
i.e. because of idiosyncratic elements, or {\it domino effect\/}, i.e.
because of failure or other agents having credit or commercial links.

In this paper we performed a first analysis of relationships of credit
between Japanese quoted firms and banks. We focus on the problem of
multiple relationships in Japan, in order to study how the typical
Japanese financial conglomerates (the {\it keiretsu\/}) influence the
network topology of the underlying architecture of credit
relationships. Notwithstanding the behavior of firms and banks is
highly heterogeneous, one may observe that firms with a large demand
for credit have multiple links (in agreement with
\citeauthor{ogawa2007jfp},~\citeyear{ogawa2007jfp}), because of risk
sharing on the part of the banks. The analysis of the MST (minimum
spanning tree) of the co-financing banks points out the presence of a
hierarchical structure of the channels of credit, with big hubs (the
largest Japanese banks) and several branches (smaller banks). These
branches have a strong geographical characterization, indicating the
presence of regional clusters in the system of the Japanese credit
market (the presence of geographical clusters is also evident as
regards the Italian market:
\citeauthor{masi2007bft},~\citeyear{masi2007bft}).

To conclude, we point out that: (i) a backbone of the credit channel
emerges, where some links play a crucial role; (ii) big banks
privilege long-term contracts; the ``minimum spanning trees'' (iii)
disclose a highly hierarchical backbone, where the central positions
are occupied by the largest banks, and emphasize (iv) a strong
geographical characterization, while (v) the clusters of firms do not
have specific common properties. Moreover, (vi) while larger firms
have multiple lending in large, (vii) the demand for credit (long vs.
short term debt and multi-credit lines) of firms with similar sizes is
very heterogeneous.

\section*{Acknowledgements}

We would like to thank the Nikkei Media Marketing, Inc.~for providing
useful information on the dataset. GDM.~acknowledges H.~Aoyama and
W.~Souma for the fruitful collaboration during the visit in Kyoto.


\newpage

\ifx\undefined\bysame
\newcommand{\bysame}{\leavevmode\hbox to\leftmargin{\hrulefill\,\,}}
\fi

\end{document}